\documentclass[conference,letter,10pt]{IEEEtran}

\usepackage{amsmath,epsfig,cite,amsfonts,amssymb,psfrag,subfig}

\def\no{\nonumber}

\newtheorem{theorem}{Theorem}
\newtheorem{pro}{Proposition}
\newtheorem{col}{Corollary}

\newtheorem{rem}{Remark}
\newtheorem{defi}{Definition}

\begin{document}

\title{Key Agreement over an Interference Channel with Noiseless Feedback: Achievable Region \& Distributed Allocation}
\vspace{-1.7cm}
\author{\IEEEauthorblockN{Somayeh Salimi, Eduard A. Jorswieck, Mikael Skoglund, Panos Papadimitratos}}

\maketitle
\vspace{-1.7cm}

\begin{abstract}
Secret key establishment leveraging the physical layer as a source of common randomness has been investigated in a range of settings. We investigate the problem of establishing, in an information-theoretic sense, a secret key between a user and a base-station (BS) (more generally, part of a wireless infrastructure), but for two such user-BS pairs attempting the key establishment simultaneously. The challenge in this novel setting lies in that a user can eavesdrop another BS-user communications. It is thus paramount to ensure the two keys are established with no leakage to the other user, in spite the interference across neighboring cells. We model the system with BS-user communication through an interference channel and user-BS communication through a public channel. We find the region including achievable secret key rates for the general case that the interference channel (IC) is discrete and memoryless. Our results are examined for a Gaussian IC. In this setup, we investigate the performance of different transmission schemes for power allocation. The chosen transmission scheme by each BS essentially affects the secret key rate of the other BS-user. Assuming base stations are trustworthy but that they seek to maximize the corresponding secret key rate, a game-theoretic setting arises to analyze the interaction between the base stations. We model our key agreement scenario in normal form for different power allocation schemes to understand performance without cooperation. Numerical simulations illustrate the inefficiency of the Nash equilibrium outcome and motivate further research on cooperative or coordinated schemes.
\end{abstract}
\let\thefootnote\relax\footnote{Somayeh Salimi, Mikael Skoglund and Panos Papadimitratos are with ACCESS Linnaeus Center, School of Electrical Engineering, KTH Royal Institute of Technology, Stockholm, Sweden, {emails: somayen@kth.se, skoglund@kth.se, papadim@kth.se}.

Eduard A. Jorswieck is with Communications Laboratory, TU Dresden, Germanye, {email: Eduard.Jorswieck@tu-dresden.de}.
 }
%
\vspace{-.1cm}

\section{Introduction}
Secret key agreement by exploiting the physical layer common randomness is a promising approach that can complement security architectures, providing with shared secret keys. The shared keys at the physical layer could be passed to the upper layers to be used for different security purposes e.g., confidentiality, authentication and integrity. The problem of secret key sharing at the physical layer has been investigated in different scenarios \cite{ITsecbook}. Secret key agreement was considered between two users in the presence of an external eavesdropper in \cite{Ahlswede} and \cite{Maurer} in which the users had access to common randomness deduced from a broadcast channel and communicated over a public channel. In some other works, the basic broadcast channel is replaced with the other basic channels which are building blocks of wireless networks. \cite{mimo},\cite{fading} and \cite{tw-bloch} consider key agreement between two users in the presence of an external eavesdropper where the common randomness arises from the fading and two-way channels. Sharing secret keys over a generalized multiple access channel is considered in \cite{salimi-mac-forward}, \cite{salimi-mac-backward}, \cite{salimi-sc+ch} and \cite{salimijsac}, in which two users intend to share secret keys with a base station hidden from each other. In these works, there is no external eavesdropper and the legitimate users are the potential eavesdroppers of each other's secret key.

In this paper, we investigate the problem of key sharing in a new scenario depicted in Fig. 1, in which BS1 communicates with User 1 while the unintended signal from BS2 reveals some information about the communications between BS2 and User 2 to User 1. Symmetrically, BS2 communicates with User 2 while the unintended signal from BS1 reveals some information about the communications between BS1 and User 1 to User 2. We assume users are honest but curious, i.e., they do not intend to spoil each other's signals, but try to obtain information about each other's communications as much as possible. The base stations are assumed to be honest and non-curious. We establish physical layer key sharing in the described scenario. According to Fig. 1, User 1 and BS1 agree on a key that is kept secret from User 2 and simultaneously, User 2 and BS2 agree on a key hidden from User 1. We model the system with an interference channel for downlink transmission from the base stations to the users. Then, we model the users communication to the base stations (uplink) as a noiseless public channel that could be eavesdropped by anyone including the neighbor cell user(s). In our scenario, first the interference channel of the downlink is used as a source of common randomness and then, the uplink public channel is used. As a result of the BS-user interaction over the interference (downlink) and public (uplink) channels, $K_{i}$ is shared between User $i$ and Base Station $i$ for $i=1,2$ according to Fig 2.

The suggested scenario is applicable to a variety of existing and upcoming technologies in the broad context of 5G, in which \emph{spectrum sharing} is used for efficient resource utilization. Due to spectrum sharing, the users near to the border of a cell suffer from inter-cell interference from the neighboring cells base stations. This not only interferes with the communications between a user and the corresponding base station, but it also results in a security challenge, i.e., information leakage. Our described key sharing scenario provides confidential communication between each user and the corresponding base station. If the base stations belong to the same operator, which already has a security architecture, then the physical layer key sharing strengthens the existing architecture. Otherwise, they can be simply used as a solution for confidentiality. Our scenario is not limited to infrastructure-based networks (e.g., cellular or mesh networks). It can also capture a general network setup with two pairs of communicating nodes (e.g., ad-hoc) under the assumption that two of the nodes are curious and two are not. It should be noted that secret key agreement over an interference channel has not been considered yet and thereby, our key agreement scheme is a novel scheme which can not be covered in the previous schemes of key agreement. Our main contributions are:
\vspace{-.1cm}

\begin{enumerate}
  \item deriving an inner bound on the secret key capacity region for the discrete memoryless setup,
  \item introducing and comparing different strategies of distributed power allocation in the Gaussian setup,
  \item non-cooperative game theoretic analysis of the distributed power allocation strategies.
\end{enumerate}


More specifically, we look for the achievable rates of key pair $(K_{1},K_{2})$. We derive an inner bound of the secret key capacity region. In the achievability scheme, two-step key generation is used in which a part of the key between each BS-user pair is established using wiretap codebooks through the downlink interference channel. The other part of the key is established through the uplink public channel exploiting secret sharing codebooks. We show that when the public channel is not used, our key sharing scenario is reduced to the secure message transmission though the interference channel which is considered in \cite{bc+if}. Then, we investigate the case of a Gaussian interference channel and numerically analyze our results. In the Gaussian case, we derive several rate regions for different transmission strategies and power allocation schemes. It is possible to have either a unidirectional or a bidirectional communication for the secret key establishment between each BS-user pair. We consider two main strategies. In a pure strategy, each BS allocates the whole available power to agree on a key either over the downlink or over the uplink. In a mixed strategy, each BS allocates a part of the available power to agree on a key over the downlink interference channel and the other part of the available power is utilized to share a key over the uplink public channel. Two power allocation schemes are used for mixed strategies; time sharing and artificial noise. All the key rate regions are compared through a numerical example.

Non-cooperative game theory proved successful for analyzing the behavior of non-cooperating links. In \cite{ifgame1}, the achievable rates of the peaceful interference channel were analyzed in a game theory framework. It was shown that the Nash equilibrium (NE) is in general far form the Pareto boundary. Here, we consider the interaction between the two BS-user pairs by a non-cooperative game. Note that there is no cooperation between the base stations and each BS-user pair tries to maximize its corresponding secret key rate. In fact, due to the inherent interference in our model, the chosen strategy by each BS affects the rate of the other BS secret key. To analyze such a bilateral effect, we exploit game theory in a non-cooperative framework in which each BS chooses its strategy independently of the other BS strategy. Obviously, the two BS-user pairs have conflicting interests not only in terms of interference to each other, but also in terms of information leakage. We define the utility functions as the respective secret key rates achieved by pure strategies and artificial noise. We show that the NE can only be achieved with pure strategies. Finally, conditions on the channel realizations and the operating SNR are derived under which only one or multiple of these strategies are NE. Numerical simulations show that in general these NE are inefficient.

As an extension of the described scenario, more than two BS-user pairs can be considered. For $K$ BS-user pairs, the downlink is modeled with a $K$-user interfere channel in which, Interference Alignment technique \cite{DOF} can be utilized for $K\geq3$. The uplink can be modeled with a public channel or in a more realistic setup, with another $K$-user interfere channel.

The rest of the paper is organized as follows: in Section II, the proposed key sharing setup is described. An inner bound on the secret key capacity region is given in Section III. The Gaussian interference channel and the corresponding rate regions are presented in Sections IV. The game analysis is given in Section V. The paper is concluded in Section VI. The proofs are given in appendices.

\begin{figure}
\centering
\includegraphics[width=7cm]{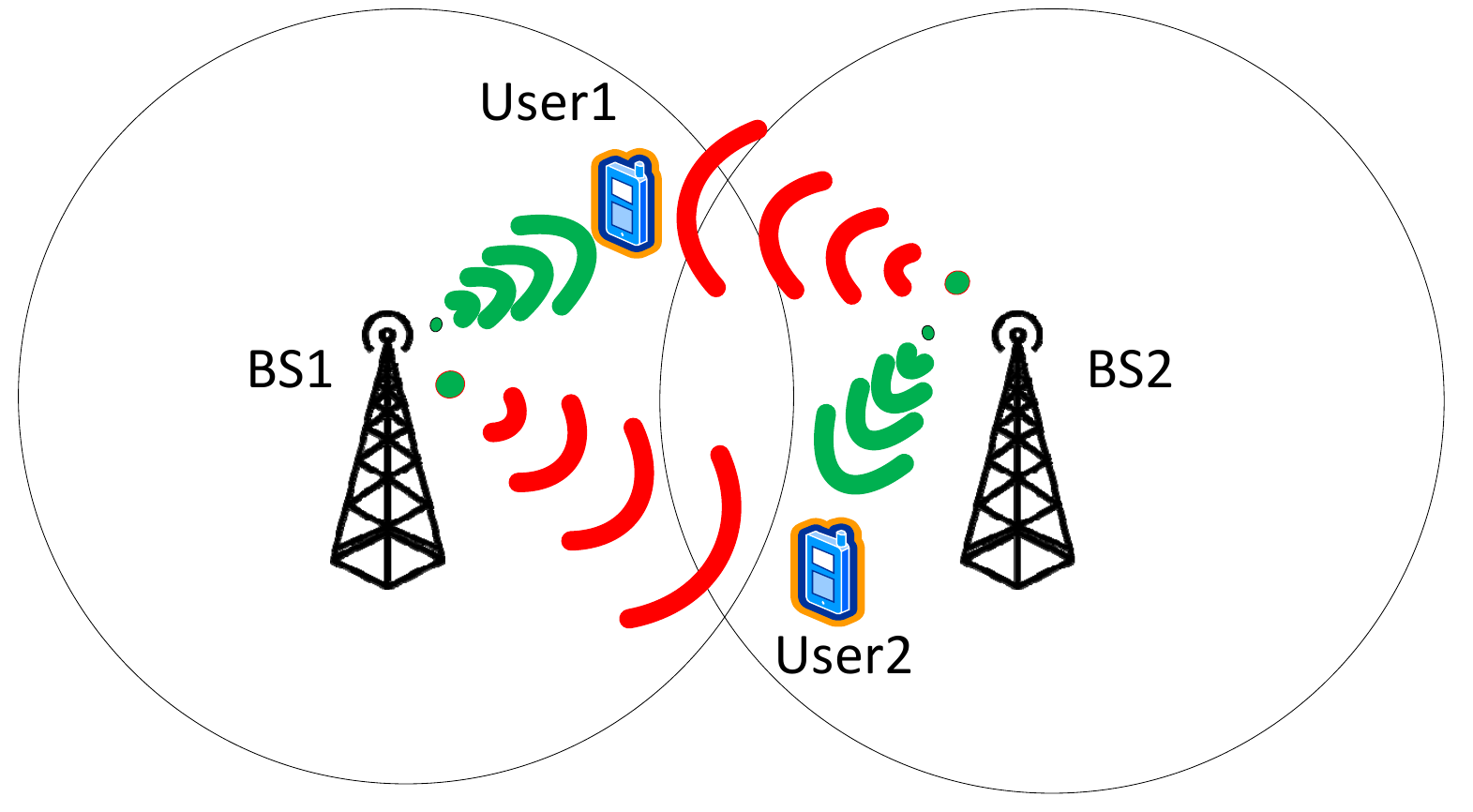}
\caption{\footnotesize{Confidentiality compromise due to inter-cell interference}
}
\vspace{-.5cm}
\end{figure}

\section{Secret Key Agreement Setup}
\noindent We assume an Interference Channel (IC) with probability distribution $P_{Y_{1},Y_{2}|X_{1},X_{2}}$, in which BS1 and BS2 govern the channel inputs $X_{1}$ and $X_{2}$ and the outputs $Y_{1}$ and $Y_{2}$ are received by Users 1 and 2, respectively. A noiseless public channel of unlimited capacity from the users to the base stations has the role of an insecure feedback channel. First, the base stations communicate with the users through the IC and then, the users communicate with the base stations over the public channel. It is assumed that the base stations make $n$ uses of the IC, and then, the users use the public channel once. In the rest of the paper, ``forward direction" is referred to as the direction from the base stations to the users, while ``backward direction" is referred to as the direction from the users to the base stations. Using the IC in the forward direction and the public channel in the backward direction, each BS-user seeks to agree on a key while keeping it concealed from the other user(s). In the following, the detailed definition of the key sharing setup as shown in Fig.2 is given.


\noindent{\emph{Step 1) $n$ uses of the IC in the forward direction}}: For $i=1,2,...,n$, BS1 and BS2 send $X_{1,i}$ and $X_{2,i}$ as the $i$th inputs of the interference channel. Subsequently channel outputs $Y_{1,i}$ and $Y_{2,i}$ are observed by Users 1 and 2, respectively.


\noindent{\emph{Step 2) Use of the noiseless feedback in the backward direction}}: Users 1 and 2, respectively, generate $F_{1}$ and $F_{2}$ as stochastic functions of $Y_{1}^{n}$ and $Y_{2}^{n}$ and send them over the public channel to the respective base stations.

After these two steps, keys $K_{\!1}$ and $K_{\!2}$ are generated by Users 1 and 2 as stochastic functions of $Y_{1}^{n}$ and $Y_{2}^{n}$, respectively. After receiving $(F_{1},F_{2})$ over the public channel, $\hat{K}_{\!1}$ and $\hat{K}_{\!2}$ are generated by BS1 and BS2 as stochastic functions of $(X_{1}^{n},F_{1},F_{2})$ and $(X_{2}^{n},F_{1},F_{2})$, respectively. Finally, $K_{1}$ is shared between BS1 and User 1 and $K_{2}$ is shared between BS2 and User 2.

\begin{rem}
At Step 2 of the key sharing setup, $F_{1}$ is actually generated by User 1 to be used by BS1, but since it is sent over the public channel, it could be in general used by BS2 for key generation. The same holds for $F_{2}$. Furthermore, we assume key $K_{1}$ as the shared key between the first BS-user pair. Since $K_{1}$ and $\hat{K}_{1}$ are the same with a high probability (as (\ref{eq2}) in Definition 1), both of them can be considered as the shared key. The same argument is valid for $K_{2}$ and $\hat{K}_{2}$.
\end{rem}

All the above keys take values in some finite sets. Now, we state the conditions that should be met in the described secret key sharing framework.

\begin{defi}
In the proposed setup, $(R_{1},R_{2})$ is an achievable key rate pair if for every $\varepsilon>0$ and sufficiently large $n$, there exists a secret key sharing code such that:
\noindent \begin{align}
&{{\tfrac{1}{n}}H(K_{1})>R_{1}-\varepsilon,{\tfrac{1}{n}}H(K_{2})>R_{2}-\varepsilon}\label{eq1}
\\&{\Pr\{K_{i}\ne \hat{K}_{i}\}<\varepsilon, \ \ \ \ \ i=1,2}\label{eq2}
\\&{{\tfrac{1}{n}}I(K_{1};K_{2},Y_{2}^{n},F_{1},F_{2})<\varepsilon}\label{eq3}
\\&{{\tfrac{1}{n}}I(K_{2};K_{1},Y_{1}^{n},F_{1},F_{2})<\varepsilon}\label{eq4}
\end{align}

\normalsize{
Equation (1) means that $R_{1}$ is the rate of the shared key between BS1 and User 1 and $R_{2}$ is the rate of the shared key between BS2 and User 2. Equation (2) means that each user and the corresponding base station generate a common key with small probability of error. Equations (3) and (4) mean that each user effectively has no information about the secret key of the other user. This refers to the weak notion of information theoretic security in which the rate, not the total amount of leaked information is negligible \cite{weeksec}.}
\end{defi}

\vspace{.15cm}
\begin{defi}
The region containing all the achievable key rate pairs $(R_{1},R_{2})$ is the secret key capacity region.
\end{defi}

\begin{figure}
\centering
\includegraphics[width=6cm]{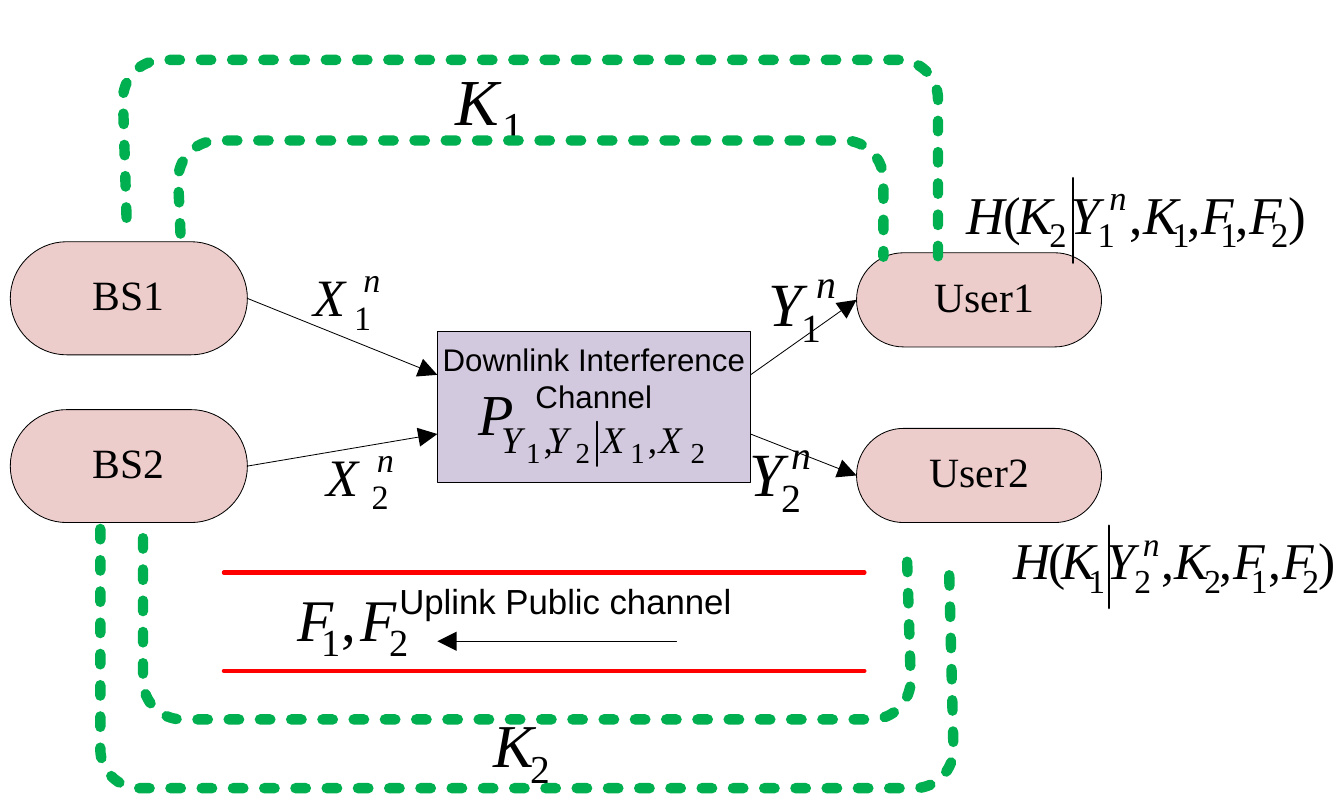}
\caption{\footnotesize{Secret key sharing over an interference channel using noiseless public channel}
}
\vspace{-.5cm}
\label{fig_sim}
\end{figure}

\section{Main Result}
\noindent We derive an inner bound on the secret key capacity region of our scheme when the interference channel is discrete memoryless.
\begin{theorem}
In the key sharing setup described in Section II, all rate pairs in the closure of the convex hull of the set of all pairs {$(R_{1},R_{2})$} {that satisfy the following conditions are achievable:}

\noindent $\begin{array}{l}
\\\hspace{-.2cm} {R_{1}\ge 0,R_{2} \ge 0,}
\vspace{.15cm}
\\\hspace{-.2cm}{R_{1}\le\![I(V_{1f};Y_{1})\!-\!I(V_{1f};\!Y_{2}|V_{2f})]^{+}+}
\\{\ \ \ \ \ \ \ \ \ [I(V_{1b};X_{1}|V_{1f})\!-\!I(V_{1b};\!Y_{2},\!V_{2f}|V_{1f})]^{+}}
\vspace{.15cm}
\\\hspace{-.2cm}{R_{2}\le\![I(V_{2f};Y_{2})\!-\!I(V_{2f};\!Y_{1},|V_{1f})]^{+}+}
\\{\ \ \ \ \ \ \ \ \ [I(V_{2b};X_{2}|V_{2f})\!-\!I(V_{2b};\!Y_{1},\!V_{1f}|V_{2f})]^{+}}
\end{array}$
\vspace{.2cm}

\noindent for random variables taking values in finite sets according to a distribution of the form:
\vspace{-.1cm}
\[\begin{array}{l} {p(v_{1f},\!v_{2f} ,\!x_{1} ,\!x_{2} ,\!y_{1} ,\!y_{2},\!v_{1b} ,\!v_{2b})\!=\!p(\!v_{1f})p(\!v_{2f})}
\\{p(\!x_{1}|v_{1f}\!)p(\!x_{2}|v_{2f}\!)p(\!y_{1},\!y_{2}|x_{1} ,\!x_{2}\!)p(\!v_{1b}|y_{1})p(\!v_{2b}|y_{2})}. \end{array}\]
\end{theorem}

The function $[x]^{+}$ equals $x$ if $x\geq0$ and 0 if $x<0$.
\vspace{.2cm}

The achievability of the key rate region in Theorem 1 is based on two-step key sharing through the IC and the public channel. At the first step, BS1 and BS2 randomly generate independent keys $K_{1f}$ and $K_{2f}$ (subscript \emph{f} stands for forward) for sharing with User 1 and User 2, respectively. Then, they encode the keys. $V_{1f}$ and $V_{2f}$ are the auxiliary random variables relevant to keys $K_{1f}$ and $K_{2f}$, respectively. Based on these auxiliary random variables, channel inputs $X_{1}$ and $X_{2}$ are generated by BS1 and BS2 and sent though the IC in $n$ uses of it. The first terms of the bounds on $R_{1}$ and $R_{2}$ correspond to this first step in which wiretap codebooks are used. After receiving the interference channel outputs by the users, they first decode the respective keys of the first step. Then, at the second step, each of the users exploits the corresponding channel output to share another key with the intended base station where the public channel is used to send the required information to the intended base station. The second terms of the bounds on $R_{1}$ and $R_{2}$ correspond to this step. $V_{1b}$ and $V_{2b}$ are the auxiliary random variables relevant to the second step keys, $K_{1b}$ and $K_{2b}$ (subscript \emph{b} stands for backward) generated by Users 1 and 2, respectively. Secret sharing codebooks are used at the second step. The detailed proof of Theorem 1 is given in Appendix I.

\vspace{.15cm}
\begin {rem}
If the public channel is not used in our setup, our result reduces to the secrecy rate region of the interference channel as Theorem 2 in \cite{bc+if} by substituting $V_{1b}=V_{2b}=\phi$ in Theorem 1.
\end {rem}

As described earlier, the pure and the mixed strategies can be considered by each BS-user pair for the key agreement. In a pure strategy, each BS-user pair shares a key either in the forward or in the backward direction but not both. In particular, a BS-user pair can choose Forward or Backward strategy as a pure strategy. In the Forward Strategy, a BS-user pair shares a key in the forward direction, while in the Backward Strategy, key sharing is performed in the backward direction. For $i=1,2$, BS-user pair $i$ chooses the Forward Strategy (FW) by setting $V_{if}\!=\!X_{i},V_{ib}\!=\!\phi$ and chooses the Backward Strategy (BW) by setting $V_{if}=\phi,V_{ib}=Y_{i}$ in Theorem 1. Hence, four situations can occur according to the chosen strategy by each BS-user pair and the bounds on $(\!R_{1}\!,\!R_{2}\!)$ are obtained as:
\begin {align}
&{([I\!(\!X_{\!1};Y_{\!1}\!)\!-\!I(\!X_{\!1};\!Y_{2},\!X_{2}\!)]^{+}\!\!,[I\!(\!X_{2};Y_{2}\!)\!-\!I(\!X_{2};\!Y_{\!1},\!X_{\!1}\!)]^{+}\!\!) (\!FW,\!FW\!)}\no
\\&{([I\!(\!X_{\!1};Y_{\!1}\!)\!-\!I(\!X_{\!1};\!Y_{2}\!)]^{+}\!\!,[I\!(\!X_{2};Y_{2})\!-\!I(\!Y_{2};\!Y_{\!1},X_{\!1})]^{+}) \ (FW,BW)}\no
\\&{([I\!(\!X_{\!1};Y_{\!1})\!-\!I(\!Y_{\!1};\!Y_{2},X_{2}\!)]^{+}\!\!,[I\!(\!X_{2};Y_{2})\!-\!I(\!X_{2};\!Y_{\!1}\!)]^{+}\!) \ (BW,FW)}\no
\\&{([I\!(\!X_{\!1}\!;\!Y_{\!1})\!-\!I(\!Y_{\!1};\!Y_{2})]^{+}\!\!,[I\!(\!X_{2};\!Y_{2})\!-\!I(\!Y_{2};\!Y_{\!1})]^{+}\!\!)(BW,\!BW\!)}
\end {align}

The chosen strategy by each BS-user pair affects the secret key rate of the other BS-user pair. For example according to (5), the bound on $\!R_{1}\!$ in (FW,BW) is strictly greater than (FW,FW) even though the first BS-user pair has the same strategy in both of them. That (FW,BW) also results in a greater bound on $\!R_{2}\!$ compared to (FW,FW) depends on the value of $I(\!X_{2};\!Y_{1},X_{1})$ and $I(\!Y_{2};\!Y_{1},X_{1})$. We will numerically analyze these strategies in the Gaussian case in Section IV.

%
%
%
%

\section{Gaussian Interference Channel}
\noindent In this section, we consider the described key sharing setup in the Gaussian Interference Channel (GIC). The GIC input-output relationships are \cite{bc+if}:
\begin{equation}{Y_{1}=X_{1}\!+\alpha_{1}X_{2}\!+\!N_{1},\ \ \  \  Y_{2}=\alpha_{2}X_{1}\!+\!X_{2}\!+\!N_{2},}\end{equation}
\noindent where the power constraints $P_{1}$ and $P_{2}$ are applied at BS1 and BS2, respectively. $N_{1}$ and $N_{2}$ are independent zero-mean, unit-variance Gaussian noise variables. In this section, we focus on the weak interference channel, i.e., $0\leq \alpha_{1}^{2}\le 1$ and $0\leq \alpha_{2}^{2}\le1$. By the standard arguments \cite{CoverBok}, the result in Theorem 1 hold in the Gaussian case. We use different transmission strategies in the following inner bound of the secret key capacity region for the Gaussian case.

First, we consider the pure strategies. By substituting the Gaussian random variables in (5), the following corollary is deduced.

\vspace{.15cm}
\begin{col}
Using the pure strategies in the Gaussian channel, the rate regions in (7) (on the top of the next page) are achievable.
\begin{figure*}[!t]
\small{
\begin{align}\no
&{[C(\frac{P_{1}}{1\!+\!\alpha_{1}^{2}P_{2}})\!-\!C(\alpha_{2}^{2}P_{1})]^{+}\!,\![C(\frac{P_{2}}{1\!+\!\alpha_{2}^{2}P_{1}})\!-\!C(\alpha_{1}^{2}P_{2})]^{+}\ \ \ \ \ \ \ \ \ \ \ \ \ \ \ \ \ \ \ \ \ \ \ \ \ \ \ \ \ \ \ \  \ \ \ \ \ \ \ \ \ \ \ (FW,FW)}\no
\\&{[C(\frac{P_{1}}{1\!+\!\alpha_{1}^{2}P_{2}})\!-\!C(\frac{\alpha_{2}^{2}P_{1}}{1+P_{2}})]^{+}\!,\!
[C(\frac{P_{2}}{1\!+\!\alpha_{2}^{2}P_{1}})\!-\!C(\frac{\alpha_{2}^{2}\!P_{1}+\!\alpha_{1}^{2}P_{2}^{2}+\alpha_{1}^{2}\alpha_{2}^{2}P_{1}P_{2}}{1\!+\!P_{2}\!+\!\alpha_{1}^{2}P_{2}})]^{+} \ \ \ \ \ \ \ \ \ \ \ \ \ \  (FW,BW)}\no
\\&{[C(\frac{P_{1}}{1\!+\!\alpha_{1}^{2}P_{2}})\!-\!C(\frac{\alpha_{1}^{2}\!P_{2}+\!\alpha_{2}^{2}P_{1}^{2}+\alpha_{1}^{2}\alpha_{2}^{2}P_{1}P_{2}}{1\!+\!P_{1}\!+\!\alpha_{2}^{2}P_{1}})]^{+},
[C(\frac{P_{2}}{1\!+\!\alpha_{2}^{2}P_{1}})\!-\!C(\frac{\alpha_{1}^{2}P_{2}}{1+P_{1}})]^{+}  \ \ \ \ \ \ \ \ \ \ \ \ \ (BW,FW)}\no
\\&{[C\!(\frac{P_{1}}{1\!+\!\alpha_{1}^{2}P_{2}})\!-\!C\!(\frac{(\alpha_{2}P_{1}\!+\!\alpha_{1}P_{2})^2}{1\!+\!(1\!+\!\alpha_{2}^{2})P_{\!1}\!+\!(1\!+\!\alpha_{\!1}^{2})P_{2}\!+\!(1\!-\!\alpha_{\!1}\alpha_{2})^{2}P_{\!1}\!P_{2}})\!]^{+}\!,}\no
\\&{\ \ \ \ \ \ \ \ \ \ \ \ \ \ \ \ [C\!(\frac{P_{2}}{1\!+\!\alpha_{2}^{2}P_{1}})\!-\!C\!(\frac{(\alpha_{2}P_{1}\!+\!\alpha_{1}P_{2})^2}{1\!+\!(1\!+\!\alpha_{2}^{2})P_{\!1}\!+\!(1\!+\!\alpha_{\!1}^{2})P_{2}\!+\!(1\!-\!\alpha_{1}\alpha_{2})^{2}P_{\!1}\!P_{2}})\!]^{+} \ \ \ \ \ \ \ \ \ \ \ \ \ \ (BW,BW)}\label{PS}
\end{align}}
\hrulefill
\end{figure*}

\normalsize{
In the equations $C(x)=\frac{1}{2}\log(1+x)$.}
\label{col:1}
\end{col}

 It should be noted that the bounds in (FW,FW) in which none of the BS-user pairs uses the public channel can be improved by using the public channel by one or both of the BS-user pairs. In the symmetric case where $P_{1}=P_{2}$ and $\alpha_{1}=\alpha_{2}$, it can be seen that $I(X_{1};\!Y_{2},X_{2})=I(Y_{1};\!Y_{2},X_{2})$ and hence, the other pure strategies outperform (FW,FW) meaning that using the public channel is beneficial.


 Although beneficial for pure strategies, the public channel utilization can be even more efficient for mixed strategies. In the following, we consider time sharing as well as artificial noise as mixed strategies.

\vspace{.15cm}
\emph{1) Time Sharing:} Time sharing is considered in \cite{bc+if} in the secrecy rate region of the Gaussian interference channel (without public channel) where the transmission period is divided into two non-overlapping slots with time fractions $\rho_{1}\geq0$ and $\rho_{2}\geq0$, where  $\rho_{1}+\rho_{2}=1$. In slot 1 with time fraction $\rho_{1}$, BS1 sends its confidential message where BS2 is silent. In slot 2 with time fraction $\rho_{2}$, BS2 sends its confidential message while BS1 remains silent. Hence, in each slot, the channel reduces to a simple Gaussian wiretap channel \cite{bc+if}. We change the time sharing scheme in \cite{bc+if} in such a way that in time fraction $\rho_{1}$, BS2 sends a signal as well which is not intended as a confidential message but it can be used in the backward phase using the public channel. In slot 2 with time fraction $\rho_{2}$, the symmetric actions are done. Therefore, in slot 1, we set:
\begin{align}
 {V_{1f}=X_{1}=\mathcal{N}(0,\beta_{1}P_{1}),V_{2f}=\phi,X_{2}=\mathcal{N}(0,\beta_{2}P_{2}),}\no
\end{align}

\noindent and in slot 2,
\begin{align}
V_{2f}=X_{2}=\mathcal{N}(0,\beta_{2}P_{2}),V_{1f}=\phi,X_{1}=\mathcal{N}(0,\beta_{1}P_{1}),\no
\end{align}

\noindent where  $0\leq\beta_{1}\leq1,0\leq\beta_{2}\leq1$ are power-control parameters.

\noindent Then in the backward phase, we set $V_{1b}=Y_{1},V_{2b}=Y_{2}$. By substituting the auxiliary random variables in Theorem 1 as described, the corollary below is resulted.

\vspace{.15cm}
\begin{col}
Using time sharing, the key rate region in (8) (on the top of the next page) is achievable over all time fraction pairs $(\rho_{1},\rho_{2})$ and power-control parameters $0\leq\beta_{1},\beta_{2}\leq1$.
\begin{figure*}[!t]
\vspace{-.8cm}
\begin{align}\no
\\&{0\leq R_{1}\leq \rho_{1}[C(\frac{\beta_{1}P_{1}}{1\!+\!\alpha_{1}^{2}\beta_{2}P_{2}})\!-\!C(\frac{\alpha_{2}^{2}\beta_{1}P_{1}}{1+\beta_{2}P_{2}})]^{+}\!+\! \rho_{2}[C(\frac{\beta_{1}P_{1}}{1\!+\!\alpha_{1}^{2}\beta_{2}P_{2}})\!-\!C(\frac{\alpha_{1}^{2}\beta_{2}\!P_{2}+\!\alpha_{2}^{2}\beta_{1}^{2}P_{1}^{2}+\alpha_{1}^{2}\alpha_{2}^{2}\beta_{1}\beta_{2}P_{1}P_{2}}{1\!+\!\beta_{1}P_{1}\!+\!\alpha_{2}^{2}\beta_{1}P_{1}})]^{+},}\no
\vspace{.5cm}
\\&{0\leq R_{2}\leq \rho_{2}[C(\frac{\beta_{2}P_{2}}{1\!+\!\alpha_{2}^{2}\beta_{1}P_{1}})\!-\!C(\frac{\alpha_{1}^{2}\beta_{2}P_{2}}{1+\beta_{1}P_{1}})]^{+}\!+\! \rho_{1}[C(\frac{\beta_{2}P_{2}}{1\!+\!\alpha_{2}^{2}\beta_{1}P_{1}})\!-\!C(\frac{\alpha_{2}^{2}\!\beta_{1}P_{1}+\!\alpha_{1}^{2}\beta_{2}^{2}P_{2}^{2}+\alpha_{1}^{2}\alpha_{2}^{2}\beta_{1}\beta_{2}P_{1}P_{2}}{1\!+\!\beta_{2}P_{2}\!+\!\alpha_{1}^{2}\beta_{2}P_{2}})]^{+},}\label{TS}
\end{align}
\hrulefill
\end{figure*}
\label{col:2}
\end{col}
Comparing the rate region in (\ref{TS}) with (FW,BW) and (BW,FW) rate regions in (\ref{PS}) demonstrates that the time sharing strategy is a combination of (FW,BW) and (BW,FW) strategies in which power control is performed.


\vspace{.15cm}
\emph{2) Artificial Noise:} Artificial noise involves splitting of the transmission
power of one of the base stations into two parts; the first is allocated to encode the message of the corresponding base station and the second part is used as artificial noise to interfere the received signal of the respective user and, hence, protect the confidential message of the other user. Thereby this scheme allows the base stations to cooperate without exchanging their confidential messages \cite{bc+if}. We propose artificial noise which is different from the one in \cite{bc+if}. In our scheme, the part of the power dedicated to artificial noise is not wasted but it can be used as a source of secrecy generation utilizing the backward public channel. When both base stations perform artificial noise as well as power control, for $i=1,2$, the auxiliary random variables in Theorem 1 are substituted as $$X_{i}=V_{if}+A_{i},V_{ib}=Y_{i}$$ where $X_{i}=\mathcal{N}(0,\beta_{i}P_{i}),V_{if}=\mathcal{N}(0,(1-\lambda_{i})\beta_{i}P_{i}),A_{i}=\mathcal{N}(0,\lambda_{i}\beta_{i}P_{i})$. In fact, BS$i$ splits its available power into two parts. A part $((1-\lambda_{i})\beta_{i}P_{i})$ is allocated to encode the secret key of the forward direction and the other part $(\lambda_{i}\beta_{i}P_{i})$ is used to confuse its corresponding user about the secret key of the other BS-user pair. The latter part of power is simultaneously used to agree on a key in the backward direction. As a result of Theorem 1, we have the following corollary.

\vspace{.15cm}
\begin{col}Using artificial noise, the key rate region in (9) (on the top of the next page) is achievable over all power-control parameters $0\!\leq\!\beta_{1},\beta_{2}\!\leq\!1$ and the power splitting
parameters $0\!\leq\!\lambda_{1},\lambda_{2}\!\leq\!1$.
\begin{figure*}[!t]
\vspace{-.8cm}
\small{
\begin{align}\no
\\&{0\leq R_{1}\leq [C(\!\frac{(\!1\!-\!\lambda_{1}\!)\beta_{1}P_{1}}{1\!+\!\alpha_{1}^{2}\beta_{2}P_{2}\!+\!\lambda_{1}\beta_{1}P_{1}}\!)\!-\!C(\!\frac{(1\!-\!\lambda_{1}\!)\alpha_{2}^{2}\beta_{1}P_{1}}{1\!+\!\alpha_{2}^{2}\lambda_{1}\beta_{1}P_{1}\!+\!\lambda_{2}\beta_{2}P_{2}}\!)]^{+}\!\!+\![C(\!\frac{(1\!-\!\alpha_{1}\alpha_{2}\!)^{2}\lambda_{1}\lambda_{2}\beta_{1}\beta_{2}P_{1}P_{2}\!+\!\alpha_{1}^{2}\lambda_{2}\beta_{2}P_{2}\!+\!\lambda_{1}\beta_{1}P_{1}}{1\!+\!\alpha_{2}^{2}\lambda_{1}\beta_{1}P_{1}\!+\!\lambda_{2}\beta_{2}P_{2}}\!)\!-\!C(\!\alpha_{1}^{2}\beta_{2}P_{2}\!)]^{+}}\no
\\&{0\leq R_{2}\leq [C(\!\frac{(\!1\!-\!\lambda_{2}\!)\beta_{2}P_{2}}{1\!+\!\alpha_{2}^{2}\beta_{1}P_{1}\!+\!\lambda_{2}\beta_{2}P_{2}}\!)\!-\!C(\!\frac{(1\!-\!\lambda_{2}\!)\alpha_{1}^{2}\beta_{2}P_{2}}{1\!+\!\alpha_{1}^{2}\lambda_{2}\beta_{2}P_{2}\!+\!\lambda_{1}\beta_{1}P_{1}}\!)]^{+}\!\!+\![C(\!\frac{(\!1\!-\!\alpha_{1}\alpha_{2})^{2}\lambda_{1}\lambda_{2}\beta_{1}\beta_{2}P_{1}P_{2}\!+\!\alpha_{2}^{2}\lambda_{1}\beta_{1}P_{1}\!+\!\lambda_{2}\beta_{2}P_{2}}{1\!+\!\alpha_{1}^{2}\lambda_{2}\beta_{2}P_{2}\!+\!\lambda_{1}\beta_{1}P_{1}}\!)\!-\!C(\!\alpha_{2}^{2}\beta_{1}P_{1}\!)]^{+}}\label{AN}
\end{align}}
\hrulefill
\end{figure*}
\label{col:3}
\end{col}

\normalsize{
Comparing the rate region in (\ref{AN}) with the rate regions of pure strategies in (\ref{PS}) shows that performing artificial noise by both base stations results in a combination of (FW,FW) (when $\lambda_{1}=\lambda_{2}=0$), (BW,BW) (when $\lambda_{1}=\lambda_{2}=1$), (FW,BW) (when $\lambda_{1}=0,\lambda_{2}=1$) and (BW,FW) (when $\lambda_{1}=1,\lambda_{2}=0$) in which, power control is performed using the parameters $\beta_{2}$ and $\beta_{2}$ at BS1 and BS2, respectively.}

The effect of using mixed strategies compared to pure strategies is shown in Fig. 3 for values $P_{1}=P_{2}=100,\alpha_{1}=\alpha_{2}=0.2$. Furthermore, the largest secrecy rate region in \cite{bc+if} (without using public channel) which is obtained performing artificial noise along with power control is shown in this figure which demonstrates the effect of using public channel. As illustrated in Fig. 3, the rate region of Corollary 3, i.e., artificial noise derived rate region includes the other regions namely the time sharing rate region and the largest rate region among the pure strategies which is (BW,BW) in this channel setup. Regarding pure strategies, it is observed that (FW,FW) leads to the smallest rate tuple compared to the other three pure strategies. That is due to the fact that choosing the FW strategy by each BS-user pair results in smaller rate for the other pair. That is because by choosing FW strategy, the base station encodes a key for the respective user which creates more side information for eavesdropping compared to the case of BW strategy in which just a random signal is sent. This fact is reflected in Fig. 3 whereas choosing FW strategy by one pair, leads to a lower rate for the other pair with a fixed strategy.
\begin{figure}
\centering
\includegraphics[width=7cm]{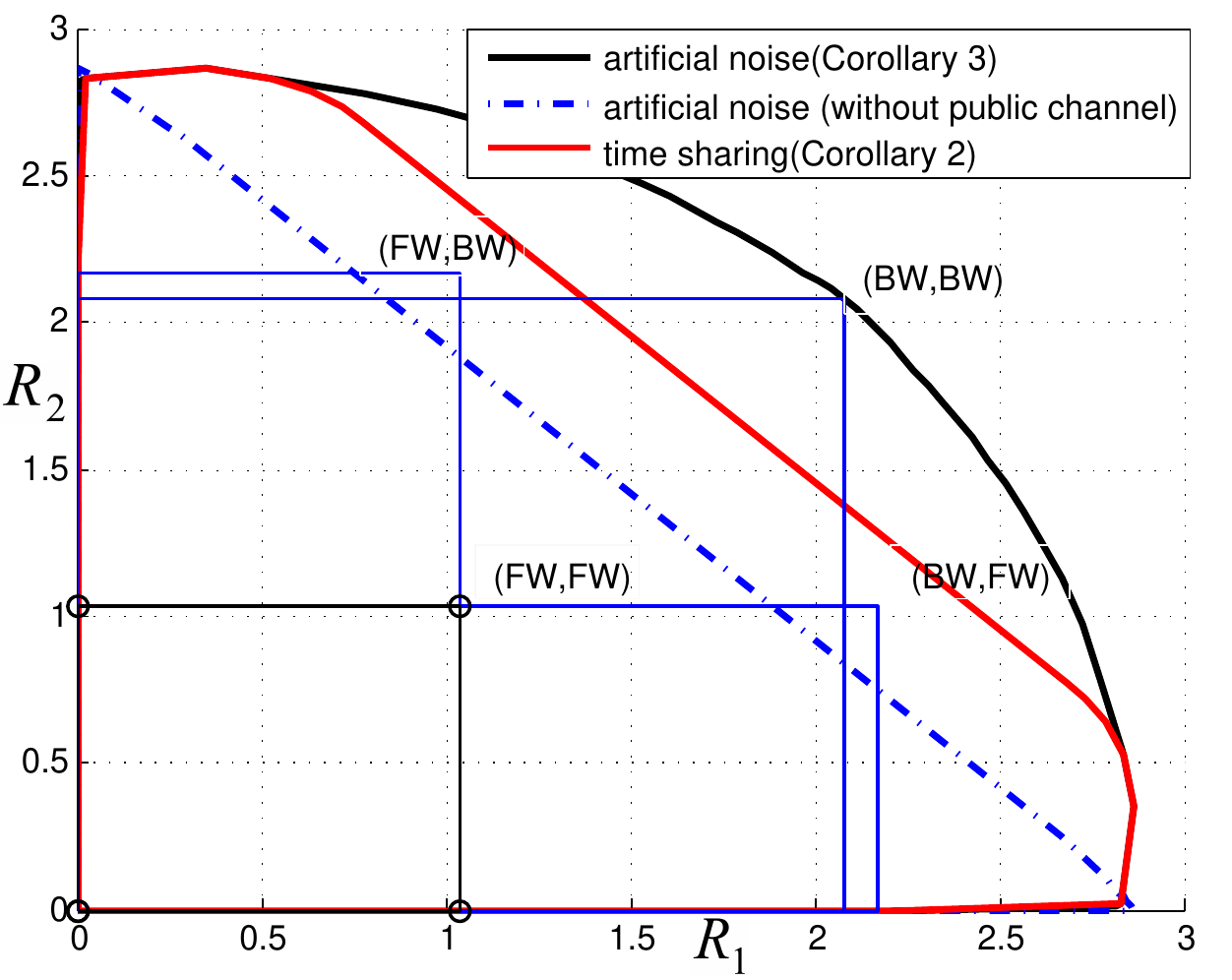}
\caption{\footnotesize{Key rate regions using different schemes}
}
\vspace{-.5cm}
\label{fig_sim}
\end{figure}

%
%
%

\section{Game analysis and Nash Equilibrium}

\noindent In this section, game analysis and Nash Equilibrium (NE) existence and uniqueness of the game in normal form are discussed. As shown earlier, the chosen strategy by each BS-user pair affects the secret key rate of the other BS-user pair. We use game theory to analyse such a two-way effect. Since each base station chooses its strategy independently of the other base station, non-cooperative game theory is exploited for such analysis. The utility functions
are defined as the respective secret key rates of the two BS-user pairs.

Note that static games in normal form can model the selfish non-cooperative behaviour of the link \cite{gametheory}. Thereby, we understand the impact of cooperation on the achievable performance. First, NE is considered in the case of pure strategies. Consider a game in strategic form $\Gamma = \left( \{ 1,2 \}, \mathcal{S}_1 \times \mathcal{S}_2, \{ R_1, R_2 \} \right)$ with players identified as BS1 and BS2, strategy space $\mathcal{S}_k$ and payoff function $R_k$ for player $k$. Note that different utility functions for the players could be used depending on the context. Here, we model the base stations as selfish.
A pure strategy pair $s_1^* \in \mathcal{S}_1$ and $s_2^* \in \mathcal{S}_2$ is called \textit{Nash equlibrium} if
\begin{eqnarray}
	R_1(s_1^*,s_2^*) \geq R_1(s_1, s_2^*) \quad \text{and} \quad R_2(s_1^*, s_2^*) \geq R_2(s_1^*, s_2) \label{eq:NE}
\end{eqnarray}
for all $s_1 \in \mathcal{S}_1$ and $s_2 \in \mathcal{S}_2$.
The strategy space for this two-player $2 \times 2$ matrix game is $\mathcal{S}_k = \{FW, BW\}$.
First, we consider arbitrary values for $\alpha_{1},\!\alpha_{2},\!P_{1}$ and $P_{2}$ in general. For $i=1,2$, if BS$i$ sets $X_{i}\!\sim\!\mathcal{N}(0,\beta_{i}P_{i})$, then the two-player $2 \times 2$ matrix game $\Gamma_1$ is completely defined using equations in (\ref{PS}).
One immediate observation from the mutual information expressions in (\ref{PS}) is that the strategy pair $(FW,FW)$ is a NE in pure strategies if and only if $\alpha_{2}^{2}\beta_{1}P_{1}\!=\!\alpha_{1}^{2}\beta_{2}P_{2}$. In order to analyze the number of NE, we reduce the parameter space and consider an interference channel in normal form with equal transmit power constraint, i.e., $\beta_1 = \beta_2 = 1$ and $P_1 = P_2 = P$. The following result characterizes the NE in the parameter space $(\alpha_1, \alpha_2) \in [0,1]^2$.

\vspace{.15cm}
\begin{pro}
For all $\alpha_1, \alpha_2 \in [0,1]^2$ there exists at least one NE in pure strategies for the game $\Gamma_1$. In medium and high SNR conditions, $P>\frac{1}{2}$, we have the following:
\begin{itemize}
  \item For $\alpha_1\!=\!\alpha_2$, there are three NE $(FW,FW),(FW,BW)$ and $(BW,FW)$.
  \item For $\alpha_1 > \alpha_2$, there is one NE $(FW, BW)$.
  \item For $\alpha_1 < \alpha_2$, there is one NE $(BW, FW)$.
\end{itemize}
	\label{prop:1}
\end{pro}
The Proof of Proposition 1 is given in Appendix II.

In Figure 4, the conditions for the NE in pure strategies are shown for $P=1$.

\begin{figure}
\centering
\includegraphics[width=7cm]{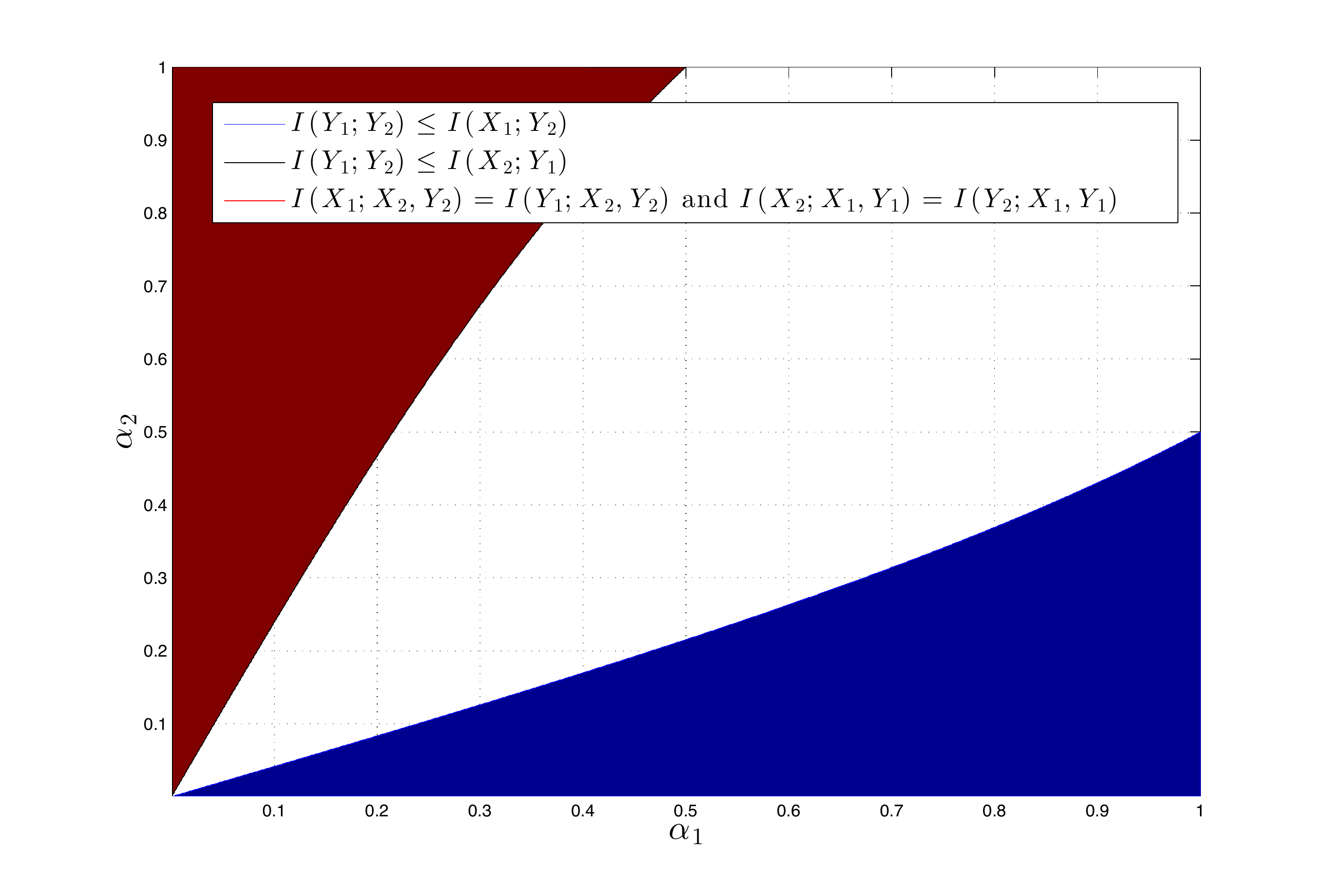}
\caption{\footnotesize{Analysis of NE in pure strategies for the game $\Gamma_1$ in strategic form for $\beta_1 = \beta_2 = 1$ and $P_1 = P_2 = 1$.}}
\vspace{-.1cm}
\end{figure}


Based on the achievable secret key rate region in Corollary \ref{col:3}, we define our next game $\Gamma_2$ in strategic form. The strategy spaces are $\mathcal{S}_1 = [0,1]$ for $\lambda_1$ and $\mathcal{S}_2 = [0,1]$ for $\lambda_2$. The utilities are the achievable secret key rates in Corollary \ref{col:3}. We set $\beta_1 = \beta_2 = 1$ because it maximizes the secret key rates of BS-user pairs. Note that game $\Gamma_2$ is a more general game in strategic form than $\Gamma_1$. However, it turns out that the best response strategies of the two players in game $\Gamma_2$ correspond to the discrete strategies from game $\Gamma_1$.

\vspace{.15cm}
\begin{pro}
 The NE of the game $\Gamma_2$ are equal to the NE of the game $\Gamma_1$ characterized in Proposition \ref{prop:1}.
 \label{prop:2}
\end{pro}

\vspace{.15cm}
 We give a sketch of the proof. Consider the best response of player one for the second game $\Gamma_2$. The secret key rate of the first BS-user pair as a function of $\lambda_1$ and $\lambda_2$ is given by
\small{
\begin{align}
 	&{R_1(\lambda_1,\lambda_2)\!=\![ \underbrace{{C ( \frac{ (1-\lambda_1) P_1}{1 + \alpha_1^2 P_2 + \lambda_1 P_1})\!-\!C( \frac{(1-\lambda_1) \alpha_2^2 P_1}{1 + \alpha_2^2 \lambda_1 P_1 + \lambda_2 P_2}) }}_{A}]^+ + \nonumber}
 \\ &\hspace{-.1cm}{[ \underbrace{{ C ( \frac{ (1 - \alpha_1 \alpha_2)^2 \lambda_1 \lambda_2 P_1 P_2 + \alpha_1^2 \lambda_2 P_2 + \lambda_1 P_1 }{1 + \alpha_2^2 \lambda_1 P_1 + \lambda_2 P_2})\!-\!C (\alpha_1^2 P_2)}}_{B}]^+} \nonumber 
 \end{align}}
 \normalsize{
 Since $[x]^+ = \max(a,0)$, the achievable secret key rate can have four different values:
 \begin{itemize}
   \item if $A\leq0,B\leq0$, then $R_1=0$
   \item if $A\geq0,B\leq0$, then $R_1=A$. In this case, the best response of the first BS-user pair to maximize $R_1$ is $\lambda_1 = 0$ (FW) irrespective of  $\lambda_2$, i.e., the strategy of the other BS-user pair.
   \item if $A\leq0,B\geq0$, then $R_1=B$. In this case, the best response of the first BS-user pair to maximize $R_1$ is $\lambda_1 = 1$ (BW) irrespective of  $\lambda_2$, i.e., the strategy of the other BS-user pair.
   \item if $A\geq0,B\geq0$, then $R_1=A+B$. In this case, the best response of the first BS-user pair is computed by calculating the first derivative of $R_1$ with respect to $\lambda_1$. It is seen that the derivative is independent of $\lambda_1$. This induces that the best response for case four depends on the strategy $\lambda_2$ and it is either $\lambda_1 = 0$ or $\lambda_1 = 1$.
 \end{itemize}

By the above arguments, in all four cases, the best response is either $0$ or $1$ and reducing the strategy space to $\lambda_1,\lambda_2 \in \{0,1\}^2$ does not reduce the number of NE.

In Figure \ref{fig_sim} it can be observed that the only pure strategy on the Pareto boundary is $(BW,BW)$ which is never a NE when $P>\frac{1}{2}$. The strategy $(FW,FW)$ is far from the Pareto boundary and also the two strategies $(FW,BW)$ and $(BW,FW)$ lie clearly within the achievable secret key region. Therefore, a coordination or cooperation mechanism is required for the two BS-user pairs to agree on a good operating point on the boundary, e.g., $(BW,BW)$.

\section{Conclusion}
\noindent
Secret key establishment in a setup including two BS-user pairs was considered in which the users were honest but curious. A combination of interference channel transmission and public channel communication was used to establish the keys. For discrete memoryless interference channel, an inner bound on the secret key capacity region was derived.
For the Gaussian interference channel, several rate regions were obtained using pure and mixed strategies. The rate regions were compared illustrating that mixed strategies including time-sharing and artificial noise outperform pure strategies. Finally, a non-cooperative game was modeled for the pure strategies and artificial noise. We showed that three of pure strategies are the only potential Nash equilibria in medium and high SNR. It was observed that the NE were inefficient and this motivates to investigate cooperative strategies as the future work. As another extension of this work, a more general setup of arbitrary number of BS-user pairs can be considered.

\normalsize{
\section*{Appendix I: Proof of Theorem 1}
\noindent We fix the distribution to be of the form as in Theorem 1. As described in Section II, the secret key sharing is established in two steps; $n$ uses of the IC and then using the public channel once by the users. In continue, we describe code construction, encoding, decoding and the security analysis.}

At the first step, wiretap codebooks are used at the base stations. BS1 and BS2 independently generate typical sequences $v_{1f}^{n}$ and $v_{2f}^{n}$, respectively, each with probability $$p(v_{1f}^{n})=\prod _{i=1}^{n}p(v_{1f,i}),p(v_{2f}^{n})=\prod _{i=1}^{n}p(v_{2f,i}).$$

\noindent The numbers of sequences $v_{1f}^{n}$ and $v_{2f}^{n}$ are $2^{n(\!r_{1\!f}+\!r'_{1\!f}\!)}$ and $2^{n(\!r_{2f}\!+r'_{2f}\!)}$, respectively, and they are labeled as:
\small{
\[\begin{array}{l}{v_{1\!f}^{n}(k_{1\!f},k'_{1\!f}),k_{1\!f}\in \mathcal K_{1\!f}\!=\!\{1,...,2^{nr_{1\!f})}\},k'_{1\!f}\!\in\!\mathcal K'_{1\!f}=\{1,...,2^{nr'_{1\!f}}\},}
\\{v_{2f}^{n}(k_{2f},\!k'_{2f}),\!k_{2f}\!\in\!\mathcal K_{2f}=\{1,...,2^{nr_{2f})}\!\},\!k'_{2f}\!\in\!\mathcal K'_{2f}\!=\!\{1,...,2^{nr'_{2f}}\!\},}\end{array}\]

\normalsize{
\noindent where \begin{equation}r'_{1\!f}\!=\!I(V_{1\!f};V_{2\!f},Y_{2})\!-\!\varepsilon',r'_{2\!f}\!=\!I(V_{2\!f};V_{1\!f},Y_{1})\!-\!\varepsilon' \label{r'} \end{equation} 

\noindent in which $\varepsilon'>0$ can be arbitrarily small.
}

\indent For the first step encoding, when a key index $k_{1f}$ is chosen by BS1, an index $k'_{1f}$ is randomly selected from $\mathcal K'_{1f}$ and then for $v_{1f}^{n}(k_{1f},k'_{1f})$, the channel input $x_{1}^{n}$ is sent according to the distribution $p(x_{1}|v_{1f})$. The same is performed by BS2.

For the first step decoding, User 1 declares error unless there exists a unique $v_{1f}^{n}(k_{1f},k'_{1f})$ such that for the received $y_{1}^{n}$, $(v_{1f}^{n},y_{1}^{n})\!\!\in\!\!A_{\varepsilon_{1}}^{n}(\!P_{V_{1f},Y_{1}}\!)$. $A_{\varepsilon_{1}}^{n}(\!P_{V_{1f},Y_{1}}\!)$ denotes a set of $\varepsilon_{1}-$jointly typical
sequences $\!(\!v_{1f}^{n},y_{1}^{n}\!)\!$ with respect to the distribution \!$p(\!v_{1f},y_{1}\!)$. User 2 acts in the same way. It can be shown that the first step decoding error probability at User $i$ is bounded as:
\[\begin{array}{l} {P_{ei,f}^{(n)} \le \varepsilon _{1} +2^{n(r_{if} +r'_{if}+\varepsilon _{1}-I(V_{if};Y_{i}))}.} \end{array}\]
%
%
If we set:

\vspace{-.3cm}
\noindent \begin{align}
&{r_{1f}+r'_{1f}<I(V_{1f};Y_{1}),}\no
\\&{r_{2f}+r'_{2f}<I(V_{2f};Y_{2}),}
  \end{align}

\normalsize{
\normalsize{\noindent then for $i=1,2$ we choose $\varepsilon_{1}=\frac{\varepsilon}{4}$ and $n$ sufficiently large such that $P_{ei,f}^{(n)}\le 2\varepsilon _{1}=\frac{\varepsilon}{2}$.}

According to the defined rates in (\ref{r'}), it can be seen that the first parts of the bounds in Theorem 1 are achievable:

\noindent \begin{align}
&{r_{1f}<I(V_{1f};Y_{1})-I(V_{1f};Y_{2}|V_{2f}),}\no
\\&{r_{2f}<I(V_{2f};Y_{2})-I(V_{2f};Y_{1}|V_{1f}),\label{forerate}}
  \end{align}

At the end of the first step, each user generates the secret key of the second step as stochastic function of the received channel output and sends the required information to the corresponding base station over the public channel. In this step, secret sharing codebook for correlated sources is used. User 1 chooses $2^{n(I(V_{1b};\!Y_{1})+\varepsilon'')}$ sequences $v_{1b}^{n}$ from $A_{\varepsilon''}^{n}(V_{1b})$ and in the symmetric way, User 2 chooses $2^{n(I(V_{2b};\!Y_{2})+\varepsilon'')}$ sequences $v_{2b}^{n}$.

\noindent These sequences are labeled using two-layered random binning as:

\small{
\noindent $\begin{array}{l}{v_{1b}^{n}(k_{1b},k'_{1b},k''_{1b}),k_{1b}\!\!\in\!\!\mathcal K_{1b}=\{1,...,2^{nr_{1b}}\},}
\\{ \ \ \ \ \ \  \ \ \ \ \ \ k'_{1b}\!\!\in\!\!\mathcal K'_{1b}=\{1,...,2^{nr'_{1b}}\},k''_{1b}\!\!\in\!\!\mathcal K''_{1b}=\{1,...,2^{nr''_{1b}}\},}
\vspace{.2cm}
\\{v_{2b}^{n}(k_{2b},k'_{2b},k''_{2b}),k_{2b}\!\!\in\!\!\mathcal K_{2b}=\{1,...,2^{nr_{2b}}\},}
\\{ \ \ \ \ \ \  \ \ \ \ \ \ k'_{2b}\!\!\in\!\!\mathcal K'_{2b}=\{1,...,2^{nr'_{2b}}\},k''_{2b}\!\!\in\!\!\mathcal K''_{2b}=\{1,...,2^{nr''_{2b}}\},} \end{array}$
}
\normalsize{

\noindent where:
}
\noindent \begin{align}&{r_{1b}\!+\!r'_{1b}\!=\!I(V_{1b};\!Y_{1}|\!Y_{2},\!V_{1\!f},\!V_{2\!f})\!+\!2\varepsilon''\!\!,}\label{rb1}
\\&{r''_{1b}\!=\!I(V_{1b};\!Y_{2},\!V_{1\!f},\!V_{2\!f})\!-\!\varepsilon''\!\!,}\label{rb''1}  \\&{r_{2b}\!+r'_{2b}=\!I(V_{2b};Y_{2}|\!Y_{1},\!V_{1\!f},\!V_{2\!f})\!+2\varepsilon'',}\label{rb2}
\\&{r''_{2b}\!=\!I(V_{2b};\!Y_{1},\!V_{1\!f},\!V_{2\!f})\!-\!\varepsilon'',}\label{rb2''} \end{align}

It is seen that for $i=1,2$, $r_{ib}+r'_{ib}+r''_{ib}\!=\!I(V_{ib};Y_{i})+\!\varepsilon''$ and hence, the sequence $v_{ib}^{n}$ can be determined with access to indices $(k_{ib},k'_{ib},k''_{ib})$.

Now, we describe the coding scheme of the second step. In this step, User 1 chooses a sequence $v_{1b}^{n}$ which is $\varepsilon''-$jointly typical with $y_{1}^{n}$. Due to the chosen rate of sequences $v_{1b}^{n}$, such sequence exists with negligible error probability. For such $v_{1b}^{n}(k_{1b},k'_{1b},k''_{1b})$, User 1 selects the respective index $k_{1b}$ as the second part of the secret key with BS1 and sends $k'_{1b}$ over the public channel. Acting in the same way, User 2 chooses a sequence $v_{2b}^{n}(k_{2b},k'_{2b},k''_{2b})$ where index $k_{2b}$ is selected to be shared with BS2 and $k'_{2b}$ is sent over the public channel.

For the second step decoding, BS$i$ decodes key $k_{ib}$ by receiving the corresponding index $k'_{ib}$ over the public channel and the information available at him, i.e., $(x_{i}^{n},v_{if}^{n})$ for $i=1,2$. Thereby BS1 decodes the sequences $v_{1b}^{n}$ if $(v_{1b}^{n}(k_{1b},k'_{1b},k''_{1b}),x_{1}^{n},v_{1f}^{n})\in A_{\varepsilon_{2}}^{n}(P_{V_{1b},X_{1},V_{1f}}),$ when such $v_{1b}^{n}$ exists and is unique. BS2 acts in the symmetric way. It can be shown that the second step decoding error probability at BS$i$ is bounded as:
\[\begin{array}{l} {P_{ei,b}^{(n)} \le \varepsilon _{2} +2^{(I(V_{ib};\!Y_{i}|X_{i} ,V_{if} )-r'_{ib}+\varepsilon _{2})}.} \end{array}\]

It we set:
\begin{align}
&{\!r'_{1b}\!>I(V_{1b};\!Y_{1}|X_{1} ,V_{1f} ),}\no
\\&{\!r'_{2b}\!>I(V_{2b} ;\!Y_{2}|X_{2},V_{2f} ),}
\end{align}
\normalsize{\noindent then for $i=1,2$ we choose $\varepsilon_{2}=\frac{\varepsilon}{4}$ and $n$ sufficiently large such that $P_{ei,b}^{(n)}\le 2\varepsilon _{2}=\frac{\varepsilon}{2}$.}

According to (\ref{rb1})-(\ref{rb2''}), the following rates are achievable for the second parts of the keys:
\begin{align}{r_{1b}<I(V_{1b};Y_{1}|\!Y_{2},\!V_{1\!f},\!V_{2\!f})-I(V_{1b};\!Y_{1}|X_{1} ,V_{1f} )}\no
\\{= I(V_{1b};X_{1}|V_{1f})\!-\!I(V_{1b};\!Y_{2},\!V_{2f}|V_{1f}),}\no \\{r_{2b}<I(V_{2b};Y_{2}|\!Y_{1},\!V_{1\!f},\!V_{2\!f})-I(V_{2b} ;\!Y_{2}|X_{2},V_{2f} )}\no
\\{=I(V_{2b};X_{2}|V_{2f})\!-\!I(V_{2b};\!Y_{1},\!V_{1f}|V_{2f}),}\label{backrate} \end{align}
The total decoding error probability at BS-user pair $i$ is bounded as:
\[\begin{array}{l} {P_{ei}^{(n)}= P_{ei,f}^{(n)}+P_{ei,b}^{(n)}\le \varepsilon .} \end{array}\]

The achievability of the secret key rates in Theorem 1 can be deduced according to (\ref{forerate}) and (\ref{backrate}).

Now, we should check the security conditions of definition 1. We give the proof of (3) and by symmetry, (4) can be deduced. Since $F_{1}=K'_{1\!b}$ and $F_{2}=K'_{2b}$, equation (3) can be rewritten as:

\noindent $\begin{array}{l}{I\!(K_{1};\!K_{2},\!Y_{2}^{n},\!F_{1},F_{2})\!\!=I\!(K_{1\!f},\!K_{1\!b};\!K_{2\!f},\!K_{2\!b},\!Y_{2}^{n},\!K'_{1\!b},\!K'_{2\!b})}
\end{array}$

As it was seen in the encoding step, $v_{2b}^{n}$ and consequently $k_{2b}$ and $k'_{2b}$ are considered as a stochastic function of $y_{2}^{n}$ and there is a Markov chain as $v_{2b}^{n}(k_{2b},k'_{2b},k''_{2b})-y_{2}^{n}-(k_{1f},k_{1b})$. Then the security condition (3) is rewritten as:

\noindent $\begin{array}{l}{I\!(K_{1\!f},\!K_{1\!b};\!K_{2\!f},\!K_{2\!b},\!Y_{2}^{n},\!K'_{1\!b},\!K'_{2\!b})=I\!(K_{1\!f},\!K_{1\!b};\!K_{2\!f},\!Y_{2}^{n},\!K'_{1\!b})}
\end{array}$

We have:

\noindent \hspace{-.2cm}$\begin{array}{l}{I\!(K_{1\!f},\!K_{1\!b};\!K_{2\!f},\!Y_{2}^{n},\!K'_{1\!b})\!\!=\!\!\underbrace{I\!(K_{1\!f};\!K_{2\!f},\!Y_{2}^{n})}_{A}\!+\!\underbrace{I\!(K_{1\!f};\!K'_{1\!b}|\!K_{2\!f},\!Y_{2}^{n})}_{B}}
\\{\!+\underbrace{I\!(K_{1\!b};\!K_{2\!f},\!Y_{2}^{n},\!K'_{1\!b}|K_{1\!f})}_{C}\!}
\end{array}$
\vspace{-.2cm}

\normalsize{
\noindent We analyze the three terms separately.

\indent For term $A$, we have:}

\small{
\[\begin{array}{l} {I\!(K_{1\!f};\!K_{2\!f},\!Y_{2}^{n})}
\\{\mathop{\le }\limits^{(a)} I(K_{1f} ;V_{2f}^{n} ,Y_{2}^{n})}
\vspace{.1cm}
\\{= H(K_{1f})-H(K_{1f},V_{1f}^{n}|V_{2f}^{n},Y_{2}^{n})+H(V_{1f}^{n}|K_{1f},V_{2f}^{n},Y_{2}^{n})}
\\{\mathop{=}\limits^{(b)}H(K_{1f})-H(V_{1f}^{n}|V_{2f}^{n},Y_{2}^{n})+H(V_{1f}^{n}|K_{1f},V_{2f}^{n},Y_{2}^{n})}
\\{\mathop{\le}\limits^{(c)}H(K_{1f})-H(V_{1f}^{n}|V_{2f}^{n},Y_{2}^{n})+n\varepsilon _{3}}
\\{\mathop{\le}\limits^{(d)}H(K_{1f})-nH(V_{1f}|V_{2f},Y_{2})+n\varepsilon _{4}+n\varepsilon _{3}}
\\{\mathop{\le}\limits^{(e)}-nH(V_{1f}|Y_{1})+n\varepsilon _{4}+n\varepsilon _{3}}
\\{\le n\varepsilon _{4}+n\varepsilon _{3}}
\end{array}\]}
\normalsize{
\noindent In the above equations, (a) and (b) are true because $k_{1f}$ and $k_{2f}$ are indices of $v_{1f}^{n}$ and $v_{2f}^{n}$, respectively. To prove (c), the same approach as Lemma 2 in \cite{bc+if} is used to show $H(V_{1f}^{n}|K_{1f},V_{2f}^{n},Y_{2}^{n})\le n\varepsilon_{3}$. (d) can be proved by exploiting the same approach as Lemma 3 in \cite{bc+if} to show $nH(V_{1f}|V_{2f},Y_{2})\le H(V_{1f}^{n}|V_{2f}^{n},Y_{2}^{n})+n\varepsilon_{4}$. (e) is the direct consequence of the reliable decoding at User 1.

\indent For term $B$, we have:}
\small{
\[\begin{array}{l} {I\!(K_{1\!f};\!K'_{1\!b}|\!K_{2\!f},\!Y_{2}^{n})}
\\{\mathop{\le }\limits^{(a)} H(K'_{1\!b}|\!K_{2\!f},\!Y_{2}^{n})-H(K'_{1\!b}|V_{1f}^{n},V_{2f}^{n},\!Y_{2}^{n})}
\vspace{.1cm}
\\{\le H(K'_{1\!b})-H(K'_{1\!b}|V_{1f}^{n},V_{2f}^{n},\!Y_{2}^{n})}
\vspace{.1cm}
\\{= H(K'_{1\!b})-H(K_{1\!b},K'_{1\!b},K''_{1\!b}|V_{1f}^{n},V_{2f}^{n},\!Y_{2}^{n})+}
\vspace{.1cm}
\\{ \ \ \ \ \ H(K_{1\!b},K''_{1\!b}|K'_{1\!b},V_{1f}^{n},V_{2f}^{n},\!Y_{2}^{n})}
\vspace{.1cm}
\\{= H(K'_{1\!b})-H(V_{1b}^{n}|V_{1f}^{n},V_{2f}^{n},\!Y_{2}^{n})+}
\vspace{.1cm}
\\{ \ \ \ \ \ H(K_{1\!b},K''_{1\!b}|K'_{1\!b},V_{1f}^{n},V_{2f}^{n},\!Y_{2}^{n})}
\vspace{.1cm}
\\{\le H(K'_{1\!b})+H(K_{1\!b})-H(V_{1b}^{n}|V_{1f}^{n},V_{2f}^{n},\!Y_{2}^{n})+}
\vspace{.1cm}
\\{ \ \ \ \ \ H(K''_{1\!b}|K_{1\!b},K'_{1\!b},V_{1f}^{n},V_{2f}^{n},\!Y_{2}^{n})}
\\{\mathop{\le }\limits^{(b)} H(K'_{1\!b})+H(K_{1\!b})-H(V_{1b}^{n}|V_{1f}^{n},V_{2f}^{n},\!Y_{2}^{n})+n\varepsilon _{5}}
\\{\mathop{\le }\limits^{(c)} H(K'_{1\!b})+H(K_{1\!b})-nH(V_{1b}|V_{1f},V_{2f},\!Y_{2})+n\varepsilon _{6}+n\varepsilon _{5}}
\\{\mathop{=}\limits^{(d)} -nH(V_{1b}|\!Y_{1},V_{1f},V_{2f},\!Y_{2})+2n\varepsilon''+n\varepsilon _{6}+n\varepsilon _{5}}
\\{\le 2n\varepsilon''+n\varepsilon _{6}+n\varepsilon _{5}}
\end{array}\]}
\normalsize{
\noindent In the above equations, (a) is true because $k_{1f}$ and $k_{2f}$ are indices of $v_{1f}^{n}$ and $v_{2f}^{n}$, respectively. To prove (b), the same approach as Lemma 2 in \cite{bc+if} is used to show $H(K''_{1\!b}|K_{1\!b},K'_{1\!b},K_{1\!f}\!K_{2\!f},\!Y_{2}^{n})\le n\varepsilon_{5}$. (c) can be proved by exploiting the same approach as in Lemma 3 in \cite{bc+if} to show $nH(V_{1b}|V_{1f},V_{2f},\!Y_{2})\le H(V_{1b}^{n}|V_{1f}^{n},V_{2f}^{n},\!Y_{2}^{n})+n\varepsilon_{6}$. (d) is the direct result of rate definition in (\ref{rb1}).

For term $C$, we have:}
\vspace{.2cm}

\small{
\noindent $\begin{array}{l}{I\!(K_{1\!b};\!K_{2\!f},\!Y_{2}^{n},\!K'_{1\!b}|K_{1\!f})\le I\!(K_{1\!b};K_{1\!f},\!K_{2\!f},\!Y_{2}^{n},\!K'_{1\!b})}
\\{\mathop{\le}\limits^{(a)}I\!(K_{1\!b};V_{1f}^{n},V_{2f}^{n},\!Y_{2}^{n},\!K'_{1\!b})}
\vspace{.1cm}
\\{=H(K_{1\!b})-H\!(V_{1b}^{n},K_{1\!b}|V_{1f}^{n},V_{2f}^{n},\!Y_{2}^{n},\!K'_{1\!b})+}
\vspace{.1cm}
\\{  \ \ \ \ \ H\!(V_{1b}^{n}|V_{1f}^{n},V_{2f}^{n},\!Y_{2}^{n},\!K'_{1\!b},K_{1\!b})}
\\{\mathop{\le}\limits^{(b)}H(K_{1\!b})-H\!(V_{1b}^{n},K_{1\!b}|V_{1f}^{n},V_{2f}^{n},\!Y_{2}^{n},\!K'_{1\!b})+n\varepsilon _{5}}
\\{\mathop{=}\limits^{(c)}H(K_{1\!b})-H\!(V_{1b}^{n}|V_{1f}^{n},V_{2f}^{n},\!Y_{2}^{n},\!K'_{1\!b})+n\varepsilon _{5}}
\vspace{.1cm}
\\{=H(K_{1\!b})-H\!(V_{1b}^{n}|V_{1f}^{n},V_{2f}^{n},\!Y_{2}^{n})+}
\vspace{.1cm}
\\{ \ \ \ \ \ \ I\!(V_{1b}^{n};\!K'_{1\!b}|V_{1f}^{n},V_{2f}^{n},\!Y_{2}^{n})+n\varepsilon _{5}}
\vspace{.1cm}
\\{\le H(K_{1\!b})-H\!(V_{1b}^{n}|V_{1f}^{n},V_{2f}^{n},\!Y_{2}^{n})+H\!(K'_{1\!b})+n\varepsilon _{5}}
\\{\mathop{\le}\limits^{(d)} H(K_{1\!b})+H\!(K'_{1\!b})-nH\!(V_{1b}|V_{1f},V_{2f},\!Y_{2})+n\varepsilon _{6}+n\varepsilon _{5}}
\\{\mathop{\le}\limits^{(e)} -nH(V_{1b}|\!Y_{1},V_{1f},V_{2f},\!Y_{2})+2n\varepsilon''+n\varepsilon _{6}+n\varepsilon _{5}}
\vspace{.1cm}
\\{\le 2n\varepsilon''+n\varepsilon _{6}+n\varepsilon _{5}}
\end{array}$}

\normalsize{
\noindent In the above equations, (a) and (b) can be resulted from the counterpart equations in deriving term B. (c) holds since $k_{1b}$ is one of the indices of $v_{1b}^{n}$. (d) and (e) are deduced from the same arguments as in (c) and (d) in deriving term B.}

Now, the total security condition (3) is obtained as:
\begin {align}
I\!(K_{1\!f},\!K_{1\!b};\!K_{1\!f},\!Y_{2}^{n},\!K'_{1\!b})\le n(\varepsilon _{4}+\varepsilon _{3}+4\varepsilon''+2\varepsilon _{5}+2\varepsilon _{6})
\end{align}

By substituting $\varepsilon''\!=\!\frac{\varepsilon}{10}$ and $\varepsilon_{i}\!=\!\frac{\varepsilon}{10}$ for $i\!=\!3,\!..,\!6$, the security condition (3) is satisfied as:
\[I\!(K_{1\!f},\!K_{1\!b};\!K_{1\!f},\!Y_{2}^{n},\!K'_{1\!b})\le n\varepsilon.\]

As the last step of proving the achievability of the rates in Theorem 1, we should demonstrate the independence of the keys $k_{1f}$ and $k_{1b}$. When analyzing term $C$ of the security condition, we showed that:
\[\begin{array}{l}{I\!(K_{1\!b};\!K_{2\!f},\!Y_{2}^{n},\!K'_{1\!b}|K_{1\!f})\le I\!(K_{1\!b};K_{1\!f},\!K_{2\!f},\!Y_{2}^{n},\!K'_{1\!b})}
\vspace{.2cm}
\\{ \ \ \ \ \ \ \ \ \ \ \ \ \ \ \ \ \ \ \ \ \  \ \ \ \ \ \ \ \  \le 2n\varepsilon''+n\varepsilon_{5} +n\varepsilon_{6},}
\end{array}\]

\noindent and consequently:
\[\begin{array}{l}{I(K_{1b};K_{1f})\le n\varepsilon.} \end{array}\]

\noindent Hence, we have:

\vspace{.4cm}
\noindent
$\begin{array}{l} {H(K_{1f},K_{1b} )\ge H(K_{1f} )+H(K_{1b} )-n\varepsilon.} \end{array}$

\vspace{.4cm}

\noindent This completes the proof of Theorem 1.
\section*{Appendix II: Proof of Proposition 1}
In order to analyze the NE for game $\Gamma_1$, the inequalities for the NE are studied. The inequalities for $(FW,FW)$ directly lead to the condition that $\alpha_2^2 = \alpha_1^2$ for all $P$.
	The conditions for $(BW,BW$) correspond to
	\begin{align}
		&\hspace{-.2cm}{\Lambda_1(\alpha_1,\!\alpha_2)\!=\!\frac{P^2 (\alpha_2\!+\!\alpha_1)^2}{1\!+\!(1\!+\!\alpha_2^2) P\!+\! (1\!+\!\alpha_1^2) P\!+\!(1\!-\!\alpha_1 \alpha_2)^2 P^2}}\label{eq:cd_BWBW1}
 \\ &{\leq \min(\alpha_1^2,\alpha_2^2) \cdot \frac{P}{1 + P}} \label{eq:cd_BWBW2}.
	\end{align}

The function $\Lambda_1$ in (\ref{eq:cd_BWBW1}) is symmetric and monotonically increasing with $\alpha_1$ and $\alpha_2$:
	\begin{align}
		& {\frac{ \partial \Lambda_1(\alpha_1,\alpha_2)}{\partial \alpha_1} =} \no
 \\ &{\frac{ 2 P^2 (\alpha_1 + \alpha_2) (P \alpha_2^2 + P + 1)(P (1- \alpha_1 \alpha_2) + 1)}{ (P^2 \alpha_1^2 \alpha_2^2 - 2 P^2 \alpha_1 \alpha_2 + P^2 + P (\alpha_1^2 + \alpha_2) + 2 P + 1)^2} > 0} \label{eq:monotonicity} 
	\end{align}
	The function in (\ref{eq:cd_BWBW2}) is increasing in $\alpha_1$ and $\alpha_2$ as well. It follows that if the inequality in (\ref{eq:cd_BWBW2}) is not fulfilled for $\alpha_1 = \alpha_2 = 1$, then it will not be fulfilled for any smaller $\alpha_1, \alpha_2$. Therefore, (BW,BW) is never a NE for $(\alpha_1,\alpha_2) \neq (0,0)$ as long as
	\begin{eqnarray}
		\frac{4 P^2}{1 + 4P} \frac{1 +P}{P} >1 \quad \Longleftrightarrow \quad P > \frac{1}{2}.
	\end{eqnarray}
	The conditions for the case (FW,BW) are:
	\begin{eqnarray}
		\frac{\alpha_2^2 P}{1 + P} \leq \Lambda_1(\alpha_1,\alpha_2) \; \text{and} \; \alpha_2 \leq \alpha_1 \label{eq:cd_FWBW}.
	\end{eqnarray}
	Since we know from (\ref{eq:monotonicity}) that $\Lambda_1(\alpha_1,\alpha_2)$ is a monotonically increasing function in $\alpha_1$, we see that $\alpha_2 \leq \alpha_1$ implies $\frac{\alpha_2^2 P}{1 + P} \leq \Lambda_1(\alpha_1,\alpha_2)$ for $P \geq \frac{1}{2}$. Therefore, only the second condition in (\ref{eq:cd_FWBW}) is relevant. The symmetric discussion holds for the case (BW,FW).

\end{document}